\documentclass[aps,pra,twocolumn,groupedaddress]{revtex4}
\usepackage{graphicx}
\usepackage{amsmath}
\usepackage{amsfonts}
\usepackage{bm}
\usepackage{color}
\usepackage[normalem]{ulem}

\newcommand{\ket}[1]{\ensuremath{\left|{#1}\right\rangle}}
\newcommand{\bra}[1]{\ensuremath{\left\langle{#1}\right |}}

\graphicspath{{converted_graphics/}}
\begin{document}

\title{Experimental entanglement redistribution under decoherence channels }

\author{G. H. Aguilar}
\affiliation{Instituto de F\'{\i}sica, Universidade Federal do Rio de Janeiro, Caixa
Postal 68528, Rio de Janeiro, RJ 21941-972, Brazil}
\email{gabo@if.ufrj.br}
\author{A. Vald\'{e}s-Hern\'{a}ndez }
\affiliation{Instituto de F\'{\i}sica, Universidade Federal do Rio de Janeiro, Caixa
Postal 68528, Rio de Janeiro, RJ 21941-972, Brazil}
\author{L. Davidovich}
\affiliation{Instituto de F\'{\i}sica, Universidade Federal do Rio de Janeiro, Caixa
Postal 68528, Rio de Janeiro, RJ 21941-972, Brazil}
\author{S. P. Walborn}
\affiliation{Instituto de F\'{\i}sica, Universidade Federal do Rio de Janeiro, Caixa
Postal 68528, Rio de Janeiro, RJ 21941-972, Brazil}
\author{P. H. Souto Ribeiro}
\affiliation{Instituto de F\'{\i}sica, Universidade Federal do Rio de Janeiro, Caixa
Postal 68528, Rio de Janeiro, RJ 21941-972, Brazil}
\date{\today }
\begin{abstract}
When an initially entangled pair of qubits undergoes local decoherence processes, there are a number of ways in which the original entanglement can spread throughout the multipartite system  consisting of the two qubits and their environments. Here we report theoretical and experimental results regarding the dynamics of the distribution of entanglement in this system.  The experiment employs an all optical set-up in which the qubits are encoded in the polarization degrees
of freedom of two photons, and each local decoherence channel is implemented with an interferometer
that couples the polarization to the path of each photon, which acts as an environment. We monitor the dynamics and distribution of entanglement and observe the transition
from bipartite to multipartite entanglement and back, and show how these transitions are intimately related to the
sudden death and sudden birth of entanglement. The multipartite entanglement is further analyzed
in terms of 3- and 4- partite entanglement contributions, and genuine four-qubit entanglement is
observed at some points of the evolution.
\end{abstract}

\maketitle

\par
\textit{Introduction:} In  one of his landmark papers published in 1935,  Erwin Schr\"odinger characterized entanglement in terms that are closely related to the modern notion of information \cite{schrodinger35}. According to him, for an entangled system ``best possible knowledge of a whole does not include best possible knowledge of its parts". In the same vein, the process of decoherence can be attributed to the loss of information of an individual quantum system to the environment with which it gets entangled \cite{zurek03}. In the case of a composite system in an entangled state, the initial entanglement, or ``information of the whole",
can spread and distribute throughout the system and
environment in a number of ways. Depending on the initial state and on the particular type of interaction
with the environment, this redistribution of the initial entanglement can give rise to
phenomena like the Entanglement Sudden Death (ESD) \cite{yu04, almeida07, salles08}, and Entanglement Sudden
Birth (ESB) \cite{lopez08}. In this paper, we address the question of what happens with the entanglement
in the cases it disappears, and how it may eventually reappear. 
This dynamical process leads to multipartite entanglement
involving system and environment. For an initial bipartite entangled state
where one subsystem interacts with the environment, it was shown \cite{farias12b, aguilar14a} that genuine
tripartite entanglement may arise in the form of GHZ \cite{ghz} or W type \cite{wstate00} of states, 
including the environmental degrees of freedom. 
 
Here, we experimentally study the situation in which two entangled
qubits are coupled to their local environments, a system that can give rise to a 
much richer dynamics, as compared to the tripartite case  \cite{farias12b, aguilar14a}, 
allowing for the detailed study of the entanglement sudden-death and sudden-birth processes,
 and the corresponding emergence of genuine multipartite entanglement involving qubits and environment. 
 The qubits are encoded in the polarizations of two photons,
while the decoherence is implemented by optically coupling
the polarization to the spatial mode,
which plays the role of the environment. Performing quantum state tomography of the whole, system and 
 environmental degrees of freedom, 
 we observe the
  entanglement as a function of the
amount of decoherence
 applied to the system. Along this evolution, we observe the
sudden-death and sudden-birth effects, and the
redistribution of the entanglement from bipartite
to tri- and four-partite forms. We also present a
theory based on monogamy relations that provides
entanglement quantifiers for some types
of entanglement.

\textit{Monogamy inequalities and residual entanglement:} In the study of the distribution of entanglement in $N$-qubit systems, the
monogamous feature of bipartite entanglement plays a fundamental role,
restricting the way bipartite entanglement can be shared among the different
components of the multipartite system. This monogamy arises from
the following inequality \cite{osborne06} 
\begin{equation}
C_{i|j_{1}j_{2}..j_{N-1}}^{2}\geq
C_{ij_{1}}^{2}+C_{ij_{2}}^{2}+...+C_{ij_{N-1}}^{2},  \label{OV}
\end{equation}%
where $C_{A|B}^{2}$ is known as the tangle and represents a measure of the
bipartite entanglement between systems $A$ and $B$. If $A$ and $B$ represent
any two subsets of qubits then

\begin{equation}
C_{A|B}^{2}=\inf_{\{     \ket{\phi_l} \bra{\phi_l},    p_l \}} \sum_{l}2 p_{l}\left[ 1-\text{tr}\left( \rho
_{A}^{l}\right) ^{2}\right] ,  \label{eq:Cmulti}
\end{equation}%
where $\rho_A^l=\text{tr}_B\left( \ket{\phi_l} \bra{\phi_l} \right) $, and the pure states $\ket{\phi_l} $ are the possible decompositions   of $\rho_{AB}$ as $\sum_l p_l \ket{\phi_l} \bra{\phi_l}$. Note that the infimum is taken over  $\{     \ket{\phi_l} \bra{\phi_l},    p_l \}$.
 If $A$ and $B$ represent single qubits, so that $A=i,$ $B=j$, then $C_{A|B}$ is the usual
two-qubit concurrence $C_{ij}$, given by \cite{wootters98} 
\begin{equation}
C_{ij}=\max \left\{ 0,\Gamma \right\} ,  \label{eq:concu}
\end{equation}%
where $\Gamma =\sqrt{\lambda _{1}}-\sqrt{\lambda _{2}}-\sqrt{\lambda _{3}}-%
\sqrt{\lambda _{4}}$ and the $\lambda _{i}$'s are the eigenvalues (in
decreasing order) of $\rho _{ij}(\sigma _{y}\otimes \sigma _{y})\rho
_{ij}^{\ast }(\sigma _{y}\otimes \sigma _{y})$. Note that the Eq. (\ref{OV}) involves
the most unbalanced bipartition $1:N-1$. For an arbitrary bipartition $M:N-M$,
monogamy relations have not been proven in the general case. Nevertheless,
for $2N$-qubit systems and partitions of the form $2:2N-2$, the following
inequality has been derived \cite{Bai09} 
\begin{equation}
C_{ii^{^{\prime }}|j_{1}j_{1}^{\prime }...j_{N-1}j_{N-1}^{\prime
}}^{2}\!\geq \!\sum_{m=1}^{N-1}C_{ij_{m}}^{2}+C_{ij_{m}^{\prime
}}^{2}+C_{i^{\prime }j_{m}}^{2}+C_{i^{\prime }j_{m}^{\prime }}^{2},
\label{monobip}
\end{equation}%
subject to the condition that the reduced density matrix $\rho _{ii^{\prime
}}$ is a rank-two matrix.

Let us now analyze the monogamy relations in the context of a 4-qubit system
composed of two open systems, S$_{1}$ and S$_{2}$, and their respective
environments, E$_{1}$ and E$_{2}$. We assume that the initial (pure) state has the structure%
\begin{equation}
\left\vert {\Psi (0)}\right\rangle _{\text{S}_1\text{S}_2\text{E}_1\text{E}_2}=\left\vert {\psi (0)}%
\right\rangle _{\text{S}_1\text{S}_2}\left\vert {0}\right\rangle _{\text{E}_1}\left\vert {0}%
\right\rangle _{\text{E}_1},  \label{psicero}
\end{equation}%
with $\left\vert {\psi (0)}\right\rangle $ an entangled state, and that only local interactions occur between systems $(\text{S}_1,\text{S}_2)$ and their respective environments, so
that the evolved state is%
\begin{equation}
\left\vert {\Psi (t)}\right\rangle =U_{\text{S}_2\text{E}_2}(t)U_{\text{S}_1\text{E}_1}(t)\left\vert {\Psi (0)}%
\right\rangle .  \label{U}
\end{equation}%
With these assumptions, direct calculation shows that the reduced density
matrices $\rho _{\text{S}_1\text{E}_1}(t)$ and $\rho _{\text{S}_2\text{E}_2}(t)$ are rank-two matrices during the whole
evolution, hence we can resort to Eq. (\ref{monobip}) and write the
 residual entanglement corresponding to the bipartition $\text{S}_1\text{E}_1|\text{S}_2\text{E}_2$ as 
\cite{Bai09} 
\begin{equation}
R_{\text{S}_1\text{E}_1|\text{S}_2\text{E}_2}\equiv C_{\text{S}_1\text{E}_1|\text{S}_2\text{E}_2}^{2}-C_{\text{S}_2\text{E}_1}^{2}-C_{\text{S}_1\text{E}_2}^{2}-C_{\text{S}_1\text{S}_2}^{2}-C_{\text{E}_1\text{E}_2}^{2},
\label{eq:residual2}
\end{equation}%
where, as seen from Eq. (\ref{monobip}), $R_{\text{S}_1\text{E}_1|\text{S}_2\text{E}_2}$ is positive. The residual entanglement corresponding to the most unbalanced
bipartition is defined via Eq. (\ref{OV}) %
\begin{equation}
R_{i}\equiv C_{i|jkl}^{2}-C_{ij}^{2}-C_{ik}^{2}-C_{il}^{2},
\label{eq:residual1}
\end{equation}%
where $i,j,k,l=\text{S}_1, \text{S}_2, \text{E}_1$ and $\text{E}_2$. The residual entanglements indicate the presence of multipartite (rather than bipartite)
entanglement. In particular, for a pure state of three qubits, the residual
entanglement becomes independent of the choice of qubit $i$ and coincides
with the 3-tangle%
\begin{equation}
\tau _{ijk}=C_{i|jk}^{2}-C_{ij}^{2}-C_{ik}^{2},  \label{Mono3}
\end{equation}
 defined in \cite{ckw00}.
 
Since the evolution of Eq.~(\ref{U}) involves only local unitary transformations, the bipartite entanglement between systems $(\text{S}_1\text{E}_1)$ and $(\text{S}_2\text{E}_2)$
remains constant. The tangle $C_{\text{S}_1\text{E}_1|\text{S}_2\text{E}_2}^{2}$ is thus conserved along the
evolution and equals the initial entanglement $\mathcal{E}%
_{0}^{2}\equiv C_{\text{S}_1\text{E}_1|\text{S}_2\text{E}_2}^{2}(0)=C_{\text{S}_1\text{S}_2}^{2}(0).$



\begin{figure}[tbp]
\centering 
\includegraphics[width=8.7cm]{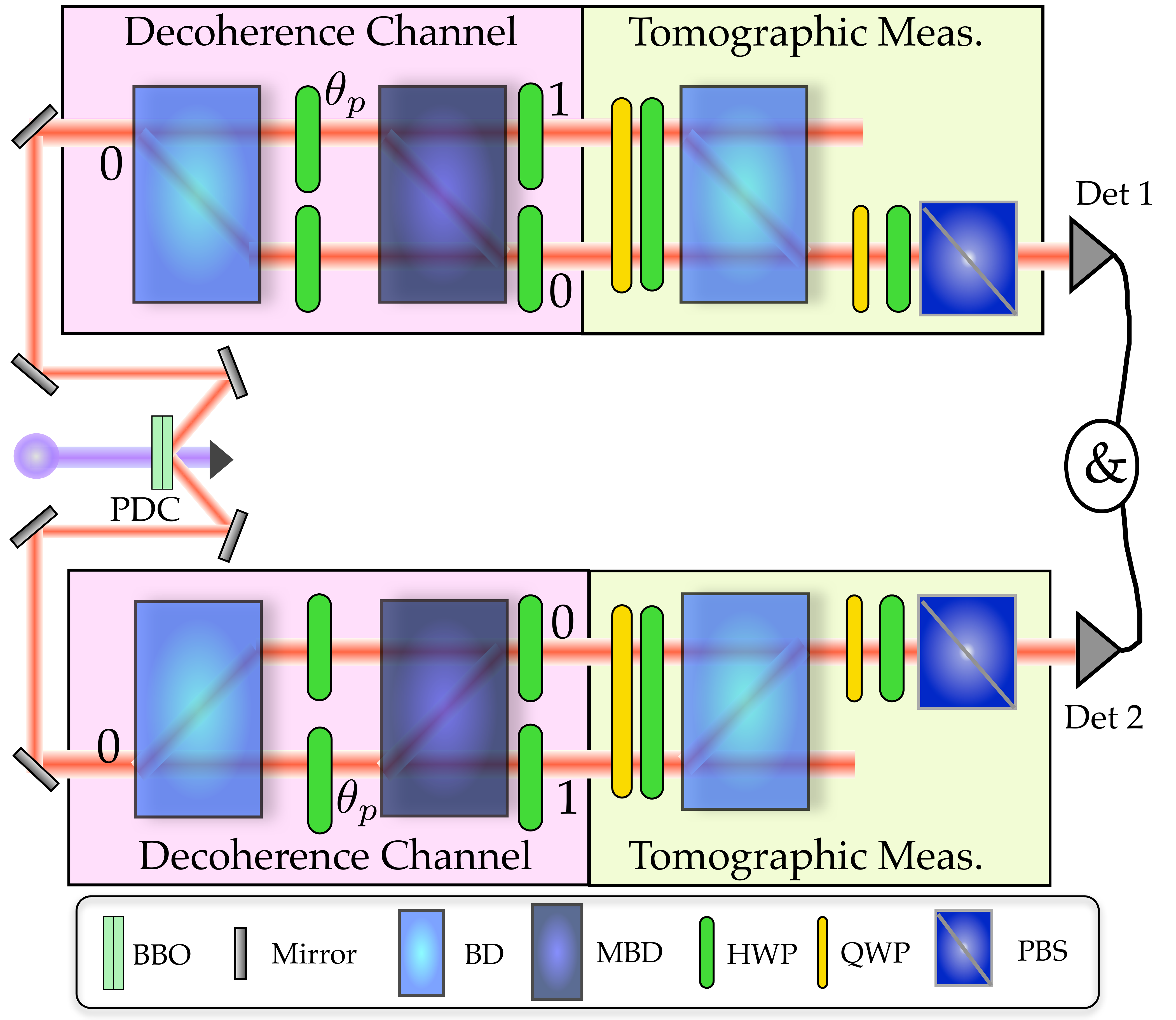}
\caption{Experimental Setup: An ultraviolet laser pumps a non-linear BBO crystal, in which two photons are produced via parametric down conversion. Both photons are directed to two separated nested interferometers. The first two interferometers (light red rectangles)  are constructed with a beam displacer (BD) and a modified beam displacer (MBD). These implements the decoherence channels as was done in \cite{farias12b, aguilar14a}.   The optical elements in the light green rectangles are used to make tomographic measurements on the polarization and path degrees of freedom simultaneously.}
\label{fig:setup}
\end{figure}

\textit{Experimental Setup:}  A sketch of the experimental setup is shown in Fig. \ref{fig:setup}.  The system qubits are encoded in the polarization of two photons, produced with
type-I spontaneous parametric down-conversion \cite{kwiat99}.  The environment is represented by the spatial mode (path) of each photon, initially in state $\ket{0}_{\text{E}}$.  Ideally, the initial two-photon state can be written as
\begin{equation}
\ket{\Psi}=(\alpha \ket{0}_{\text{S}_1}\ket{0}_{\text{S}_2}+\beta \ket{1}_{\text{S}_1}\ket{1}_{\text{S}_2})\ket{0}_{\text{E}_1}\ket{0}_{\text{E}_2},
\label{eq:est_init}
\end{equation}
where $\ket{0}_{\text{S}}$ and $\ket{1}_{\text{S}}$ are represented by horizontal and vertical polarization of the photons, respectively. Each one of the photons is directed to optical interferometers that implement
a unitary transformation that models the interaction between the polarization $\text{S}_i$ and the spatial mode $\text{E}_i$. Tracing out the environment degrees of freedom, this operation corresponds to a 
quantum channel \cite{farias12b, aguilar14a}.  A second interferometer, wave plates and a polarizing beam splitter (PBS) are used to perform
full state tomography on the polarization and path degrees of freedom of each photon.  In the following, we briefly summarize the role of each interferometer, as complete details can be found in Refs. \cite{farias12b, aguilar14a}.  
\par

The interferometers are constructed using calcite beam displacers (BD, MBD) to separate horizontal and vertical polarization components.  A half-wave plate (HWP), whose rotation angle is $\theta_p$  in Fig. \ref{fig:setup}, is used to control the amount of decoherence in this case.  As $\theta_p$ varies, the spatial mode 1 is populated according to
the quantum map:
\begin{equation}
\begin{split}
|{0}\rangle_{\text{S}}|{0}\rangle_{\text{E}}&
\rightarrow |{0}\rangle_{\text{S}}|{0}\rangle_{\text{E}},
\\
|{1}\rangle_{\text{S}}|{0}\rangle_{\text{E}}&
\rightarrow \sqrt{1-p}| {1}\rangle_{\text{S}}|{0}
\rangle_{\text{E}}+\sqrt{p}| {0}\rangle_{\text{S}}|{1}
\rangle_{\text{E}},
\end{split}
\label{eq:ad}
\end{equation}
where $p=\sin^2(2\theta_p)$. This transformation corresponds to the amplitude damping
channel (AD) when the environments are traced out. The analogy with a relaxation process can be made by associating the parameter
$p$ with time, so that $p(t=0) = 0$, and $p(t \rightarrow \infty) = 1$.
The second interferometer and the polarization optics shown in the light green rectangles of Figs. \ref{fig:setup} are used to perform complete tomography of the polarization and spatial mode using 256 different settings of the HWPs and quarter-wave plates (QWP) \cite{farias12b, aguilar14a, tomography}.  That is, we perform complete 4 qubit tomography, allowing us to reconstruct the total quadripartite state of system + environment, and to completely analyze the evolution of entanglement as a function of $p$.   

\begin{figure}[tbp]
\centering 
\includegraphics[width=8cm]{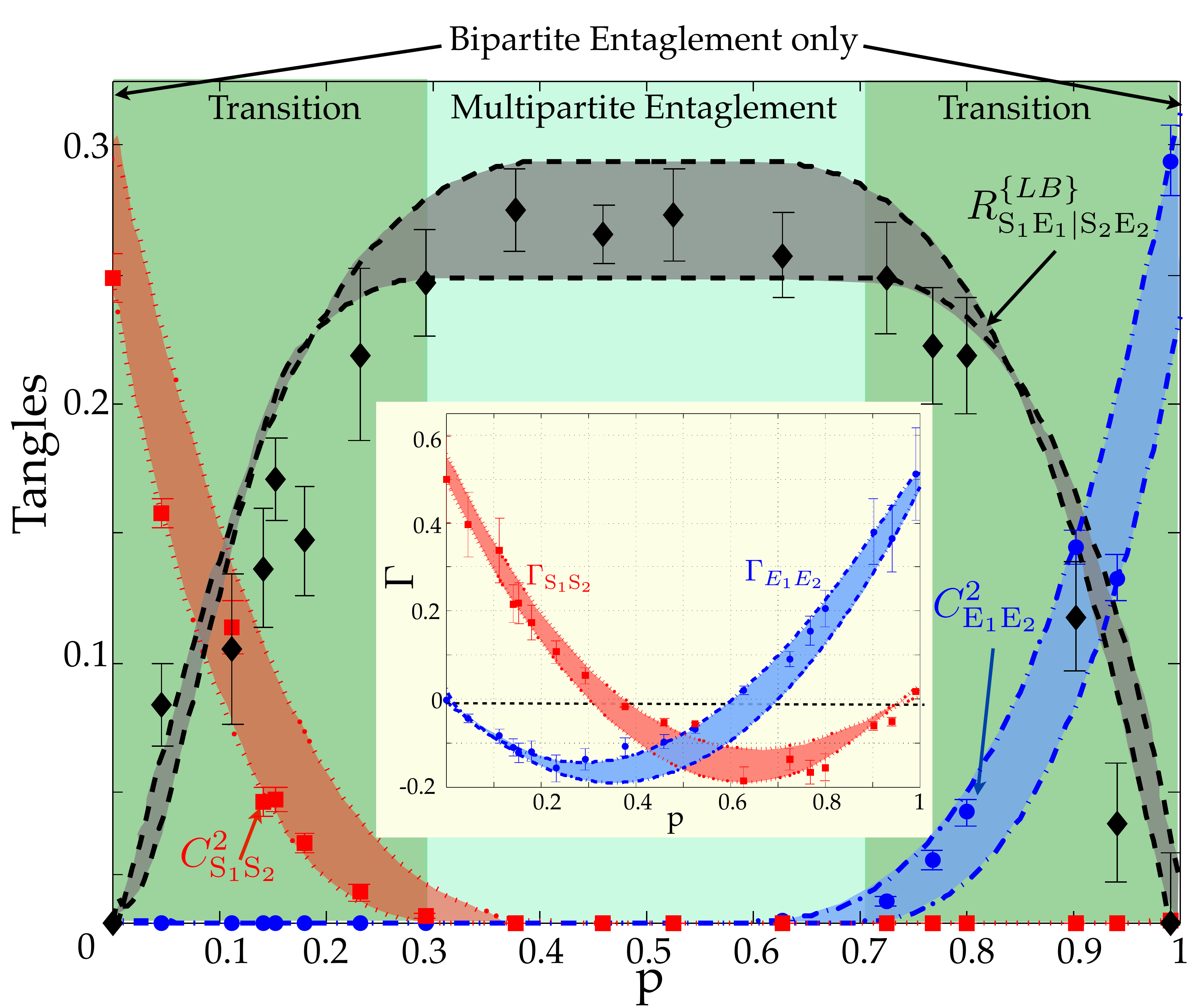}
\caption{Experimental tangles $C_{ij}^{2}$ versus $p$, the evolution
parameter. The red squares represent $C_{\text{S}_{1}\text{S}_{2}}^{2}$,
which decays monotonically until $p= 0.34 \pm 0.04$  where the sudden death of
entanglement takes place. $C_{\text{E}_{1}\text{E}_{2}}^{2}$ (blue circles)
is null for $p$ in the range [0, 0.62] where it begins to monotonically
grow, the entanglement sudden birth. Black diamonds represent the
residual entanglement of Eq. (\ref{eq:residual2}), in the Lower Bound
approximation \protect\cite{mintert07}. The areas are theoretical predictions (see the text). In the inset we show $\Gamma $,
defined in Eq. (\protect\ref{eq:concu}), where one can see that the
entanglement is actually born at $p_{ESB}$.}
\label{fig:results1}
\end{figure}

\textit{Distribution of entanglement:} We analyze the ``sudden death"
and ``sudden birth" of entanglement, showing that what happens in fact is
a redistribution of the initial entanglement between system and environment. 
For this purpose, we create photons
in the state given by Eq. (\ref{eq:est_init}) with $\alpha \simeq \sqrt{1/7}$,
 $\beta \simeq \sqrt{6/7}$, with purity 0.82 and fidelity 0.9. Both photons are sent to  the interferometers, which implement AD channels as given in Eq. (\ref{eq:ad}). We vary the evolution parameter $p_{i}$ for both channels, so that $p_{1}=p_{2}=p$. The evolved states of the complete system can be
written as 
\begin{eqnarray}
\left\vert {\Psi (p)}\right\rangle _{\text{S}_1 \text{S}_2 \text{E}_1 \text{E}_2} &=&\!\!\frac{1}{\sqrt{7}}%
\left\vert {0000}\right\rangle +\sqrt{\frac{6}{7}}[(1-p)\left\vert {1100}%
\right\rangle    \notag \\
&+&p\left\vert {0011}\right\rangle+\sqrt{p(1-p)}\left\vert {1001}\right\rangle \notag\\
&+&\sqrt{p(1-p)}\left\vert {0110}\right\rangle].
   \label{evol}
\end{eqnarray}%
Full quantum state tomography was performed and entanglement was calculated for several values of $p$, as shown in Fig. \ref{fig:results1}. We can see
that the entanglement between the systems S$_{1}$ and S$_{2}$ (red squares) decays
monotonically until the complete disappearance of entanglement (sudden death), at $p_{ESD}=0.34 \pm 0.04$. This value is close to
 the theoretical value given by $|\alpha/\beta|=\sqrt{1/6}\simeq 0.4 $. The discrepancy  can be attributed to the fact that $p_{ESD}$ is $|\alpha/\beta|$ for pure states, while the experimental states are slightly mixed. At $p_{ESB}=0.67 \pm 0.05 $ we observe the reappearance of bipartite
entanglement. However, it has now been swapped to the environment qubits E$_{1}$ and E$_{2}$ \cite%
{lopez08}. This result is close to the theoretical value $1-|\alpha/\beta| \simeq 0.6 $, obtained for pure states. In the inset of Fig. \ref{fig:results1},
we show the evolution of $\Gamma$ defined in Eq. (\ref{eq:concu}). As we
can see, $\Gamma _{\text{E}_{1}\text{E}_{2}}$, represented by blue circles, takes negative values for $%
p<p_{ESB}$, assuring that the entanglement between E$_1$ and E$_2$
(measured by the concurrence in Eq. (\ref{eq:concu})) is null before the
entanglement sudden birth. 
This corroborates the claim that the entanglement
between E$_{1}$ and E$_{2}$ is actually born at $p_{ESB}$. In the same way,
we can see that after the entanglement sudden death, $\Gamma _{\text{S}_{1}%
\text{S}_{2}}$ also takes negative values.
 The values of $C_{\text{S}_{1}\text{E}_{2}}^{2}$ and $C_{\text{S}%
_{2}\text{E}_{1}}^{2}$ are very close to zero during the entire evolution and
are not shown. Black diamonds indicate the residual entanglement $R_{\text{S}%
_{1}\text{E}_{1}|\text{S}_{2}\text{E}_{2}}$ defined in Eq. (\ref{eq:residual2}), where the values of $C_{\text{S}%
_{1}\text{E}_{1}|\text{S}_{2}\text{E}_{2}}^{2}$ were obtained in the lower-bound
approximation \cite{mintert07, farias12b}:
\begin{equation}
C_{\text{S}_{1}\text{E}_{1}|\text{S}_{2}\text{E}_{2}}^{2}\geq [C_{\text{S}_{1}\text{E}_{1}|\text{S}_{2}\text{E}_{2}}^{\{LB\}}]^{2}= 2[\text{tr}(\rho)-\text{tr}(\rho_{\text{S}_{1}\text{E}_{1}})],
\end{equation}
where $\rho=\rho_{\text{S}_1\text{S}_2 \text{E}_1 \text{E}_2}$ is the complete density matrix  and $\rho_{\text{S}_1\text{E}_1 } $ is the reduced density matrix.
The theoretical predictions for the experimental results are given by the curved shaded regions in Fig. \ref{fig:results1}, which we will now describe.Ê The curves that limit the shaded regions from below correspond to the theoretical unitary evolution of the initial ($p=0$) experimental state, while the curves that limit from above correspond to theoretical evolution of the final state ($p=1$) backwards.Ê In the absence of experimental imperfections, these curves would coincide. Ê Therefore, the shaded areas represent regions where the experimental points can be considered compatible with the theory.

For $p_{ESD}\leq p\leq p_{ESB}$, no (bipartite) qubit-qubit entanglement
contributing to Eq. (\ref{eq:residual2}) is observed, and in agreement with the theoretical result in Ref. \cite{mcnemes}, the residual
entanglement $R_{\text{S}_{1}\text{E}_{1}|\text{S}_{2}\text{E}_{2}}$ reaches its maximum value. This means that the entanglement changes its nature along
the evolution. In the beginning ($p=0$) it is given entirely by bipartite
entanglement between S$_{1}$ and S$_{2}$. In the transition interval (dark green
region) $p\in (0,p_{ESD})$ bipartite and multipartite entanglement coexist.
Between $p_{ESD}$ and $p_{ESB}$ (light green region) the entanglement is entirely multipartite, and after another
transition interval ($p\in (p_{ESB},1]$) the evolution ends up with
the same initial bipartite entanglement, but now swapped to the environments
E$_{1}$ and E$_{2}$.


\textit{Further decomposition of the multipartite entanglement:} We now  go
into more detail regarding the type of multipartite entanglement that
appears during the evolution. An explicit decomposition of $R_{\text{S}_1 \text{E}_1|\text{S}_2\text{E}_2}$ in
terms of well-identified multipartite entanglements is needed to provide
information about the distribution of entanglement.  We
first use Eqs. (\ref{eq:residual1}) (applied to $i=\text{S}_1,\text{E}_1$) and (\ref%
{eq:residual2}) to write 
\begin{equation}
\mathcal{R}_{\text{S}_1\text{E}_1}=\mathcal{E}%
_{0}^{2}+2C_{\text{S}_1\text{E}_1}^{2}-(C_{\text{S}_1|\text{S}_2 \text{E}_1 \text{E}_2}^{2}+C_{\text{E}_1|\text{S}_1\text{S}_2\text{E}_2}^{2}),  \label{conserv2}
\end{equation}%
where we have defined $\mathcal{R}_{\text{S}_1\text{E}_1}\equiv R_{\text{S}_1\text{E}_1|\text{S}_2\text{E}_2}-(R_{\text{S}_1}+R_{\text{E}_1}).$
Since $\mathcal{R}_{\text{S}_1\text{E}_1}$ is defined as a linear combination of residual
multipartite entanglements, it must be possible to express it in terms of
non-bipartite entanglement contributions only. In order to do this we first
use Eq. (\ref{U}) and observe that $\left\vert {\Psi (t)}\right\rangle 
$ can be obtained by applying $U_{\text{S}_2\text{E}_2}$ to the intermediate state $%
U_{\text{S}_1\text{E}_1}(t)\left\vert {\Psi (0)}\right\rangle .$ As is stated below Eq. (\ref{U}), the reduced density matrix $\rho _{\text{S}_2\text{E}_2}$ is a
rank-two matrix, so that at this intermediate stage the problem is that of a 3-qubit ($%
\text{S}_1$, \thinspace $\text{E}_1$ and $(\text{S}_2\text{E}_2)$) system in a pure state, and in which only $\text{S}_1$
and \thinspace $\text{E}_1$ interact. In this case one multipartite entanglement
arises, corresponding precisely to the 3-tangle in Eq. (\ref{Mono3}) with
the qubit $i$ being $\text{S}_1$, $j$ being $\text{E}_1$ and $k$ being the effective two-level system $%
(\text{S}_2\text{E}_2)$. We  denote this quantity as $\tau _{\text{S}_1\text{E}_1(\text{S}_2\text{E}_2)}$.

Now, by invoking the invariance of entanglement under local operations, it follows that $\tau
_{\text{S}_1\text{E}_1(\text{S}_2\text{E}_2)}$ remains unaffected when the transformation $U_{\text{S}_2\text{E}_2}$ is applied to
the state $U_{\text{S}_1\text{E}_1}(t)\left\vert {\Psi (0)}\right\rangle ,$ so that the final
state $\left\vert {\Psi (t)}\right\rangle $ has a multipartite entanglement $%
\tau _{\text{S}_1\text{E}_1(\text{S}_2\text{E}_2)}$ whose value is independent of $U_{\text{S}_2\text{E}_2}$. On the other hand,
as seen from Eq. (\ref{conserv2}), $\mathcal{R}_{\text{S}_1\text{E}_1}$ depends only on the
reduced density matrices $\rho _{\text{S}_1\text{E}_1},\rho _{\text{S}_1}$ and $\rho _{\text{E}_1},$ and
consequently represents a multipartite entanglement that does not depend on
the transformation $U_{\text{S}_2\text{E}_2}$ neither. Since $\tau _{\text{S}_1\text{E}_1(\text{S}_2\text{E}_2)}$ and $\mathcal{R}_{\text{S}_1\text{E}_1}$ are both independent of $U_{\text{S}_2\text{E}_2}
$ we can compute them assuming that $U_{\text{S}_2\text{E}_2}=\mathbb{I}.$ In this case $\text{E}_2$ remains in its ground state, the system $(\text{S}_1\text{S}_2\text{E}_1)$ remains in a pure
state and thus $\tau _{\text{S}_1\text{E}_1(\text{S}_2\text{E}_2)}$ is simply given by the 3-tangle $\tau _{\text{S}_1\text{S}_2\text{E}_1}$
that corresponds to the 3-qubit state calculated with equation (\ref{Mono3}). As for $%
\mathcal{R}_{\text{S}_1 \text{E}_1},$ we notice that with $U_{\text{S}_2\text{E}_2}=\mathbb{I}$ we can apply the
decomposition (\ref{Mono3}) to $C_{\text{S}_1|\text{S}_2\text{E}_1\text{E}_2}^{2}=C_{\text{S}_1|\text{S}_2\text{E}_1}^{2}$, $%
C_{\text{E}_1|\text{S}_1\text{S}_2\text{E}_2}^{2}=C_{\text{E}_1|\text{S}_1\text{S}_2}^{2},$ and to $\mathcal{E}%
_{0}^{2}=C_{\text{S}_1\text{E}_1|\text{S}_2\text{E}_2}^{2}=C_{\text{S}_2|\text{S}_1\text{E}_1}^{2},$ thus obtaining 
$\mathcal{R}_{\text{S}_1\text{E}_1}=-\tau _{\text{S}_1\text{S}_2\text{E}_1}=-\tau
_{\text{S}_1\text{E}_1(\text{S}_2\text{E}_2)}$.
From the definition of $\mathcal{R}_{\text{S}_1\text{E}_1}$ we are  led to write 
\begin{equation}
R_{\text{S}_1\text{E}_1|\text{S}_2\text{E}_2}=R_{\text{S}_1}+R_{\text{E}_1}-\tau _{\text{S}_1\text{E}_1(\text{S}_2\text{E}_2)}.  \label{r13}
\end{equation}%
Decomposing the residual entanglements $R_i$, we can rewrite $R_{S_1E_1|S_2E_2}$ in terms of
more explicit multipartitie entanglement contributions. This can be accomplished considering
that if $\rho _{\text{S}_2\text{E}_2}$ is a rank-two matrix, so that the subsystem $(\text{S}_2\text{E}_2)$ can be
considered as single qubit, and the complete system $\text{S}_1\text{E}_1(\text{S}_2\text{E}_2)$ is in a pure
state, then we can use Eq. (\ref{Mono3}) and write 
\begin{eqnarray}
C_{i|jkl}^{2} &=&C_{i|j(kl)}^{2}=C_{ij}^{2}+C_{i(kl)}^{2}+\tau _{ij(kl)} 
\notag \\
&=&C_{ij}^{2}+C_{ik}^{2}+C_{il}^{2}+\tau _{\underline{i}kl}+\tau _{ij(kl)},
\label{desco}
\end{eqnarray}%
where in the second line we used Eq. (\ref{OV}) and denoted with $%
\tau _{\underline{i}kl}$ the corresponding residual entanglement. The
underline distinguishes the reference qubit and stresses the fact that $\tau
_{\underline{i}kl}$ is not necessarily invariant under a permutation of
indices that involves $i,$ as is $\tau _{ijk}$ in Eq. (\ref{Mono3}). In
fact, the 3-partite entanglement $\tau _{\underline{i}kl}$ coincides with
the 3-tangle $\tau _{ijk}$ only when the subsystem $ikl$ is in a pure state,
that is, only when $j$ is separable from the rest of the qubits. Comparison of Eqs.  (\ref{eq:residual1}) and (\ref%
{desco}) leads to 
\begin{equation}
R_{i}=\tau _{\underline{i}kl}+\tau _{ij(kl)},  \label{Ri}
\end{equation}%
an expression that allows us to compute the 3-partite entanglement $\tau _{%
\underline{i}kl}$ of the mixed state $\rho _{ikl}$ as 
\begin{equation}
\tau _{\underline{i}kl}=C_{i|jkl}^{2}-(C_{ij}^{2}+C_{ik}^{2}+C_{il}^{2})-%
\tau _{ij(kl)}.  \label{3partite}
\end{equation}%
Equations  (\ref{r13}), (\ref{Ri}), and the analogous equation that results
from the latter by performing the substitution $1\leftrightarrow 2$, lead to%
\begin{eqnarray}
R_{\text{S}_{1}\text{E}_{1}|\text{S}_{2}\text{E}_{2}} &=&\tfrac{1}{2}[\tau _{%
\underline{\text{S}_{1}}\text{S}_{2}\text{E}_{2}}+\tau _{\underline{\text{S}%
_{2}}\text{S}_{1}\text{E}_{1}}+\tau _{\underline{\text{E}_{1}}\text{S}_{2}%
\text{E}_{2}}+
\label{res} \\
&&\tau _{\underline{\text{E}_{2}}\text{S}_{1}\text{E}_{1}}+\tau _{\text{S}_{1}\text{E}_{1}(\text{S}_{2}\text{E}_{2})}+\tau _{\text{S}%
_{2}\text{E}_{2}(\text{S}_{1}\text{E}_{1})}].  \notag
\end{eqnarray}
\begin{figure}[tbp]
\centering 
\includegraphics[width=8 cm]{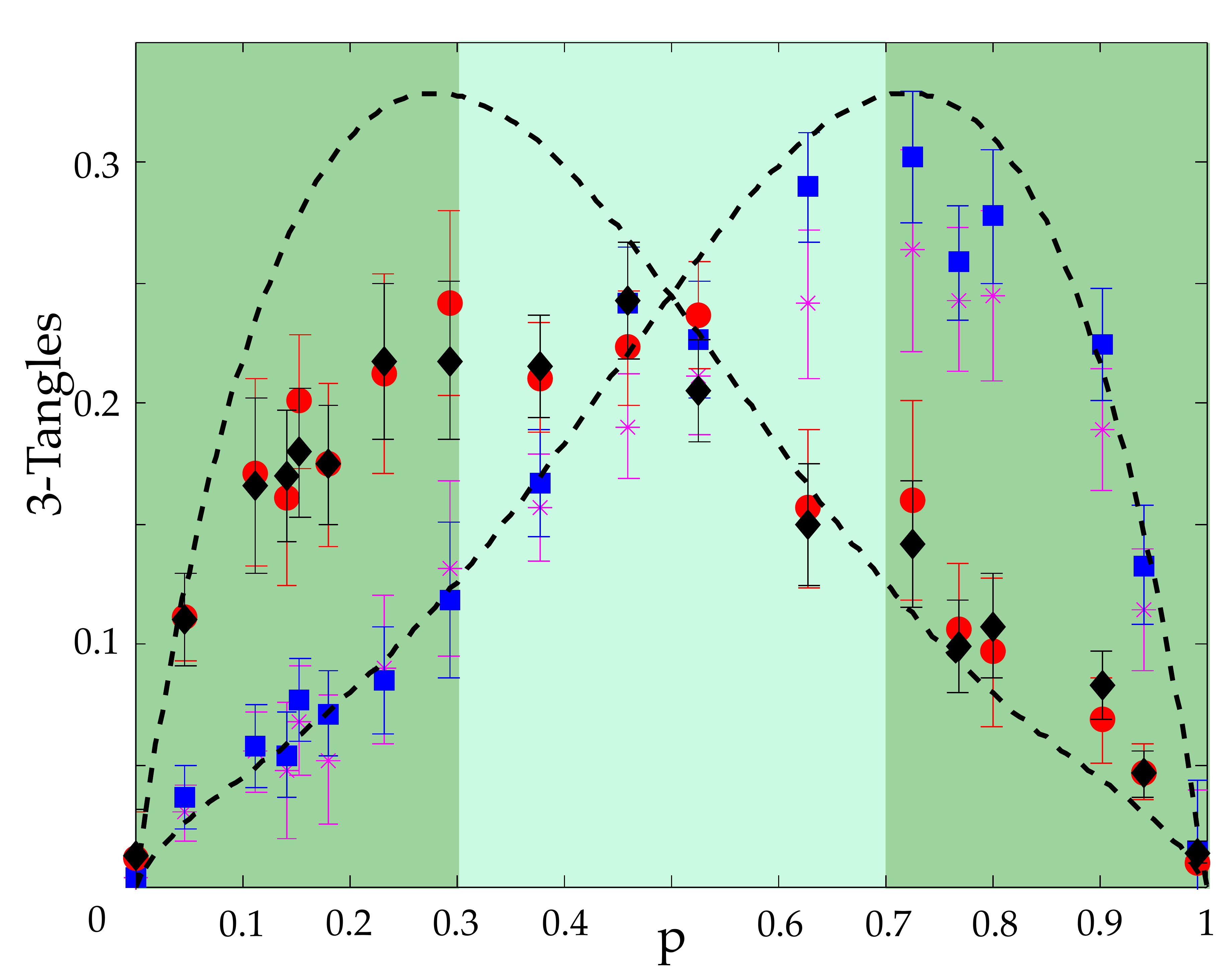}
\caption{ Three-tangles $\protect\tau _{\protect\underline{i}jk}$ in the
quasi pure approximation versus $p$. As theoretically expected, $
\tau _{\underline{S_1}S_2E_2}$ (red circles) and $\tau _{\underline{S_2}S_1E_1}$ (black diamonds) follow nearly the same evolution. The same
happens for $\protect\tau _{\protect\underline{E_1}S_2E_2}$ (blue squares) and $%
\protect\tau _{\protect\underline{E_2}S_1E_1}$ (magenta stars). The lines are theoretical prediction for pure states.}
\label{fig:results2}
\end{figure}

With these results we now analyze our experimental data  regarding the residual entanglements $R_{i}$ which, as seen from Eq. (\ref%
{Ri}), can be separated in a tripartite entanglement contribution, measured
by $\tau _{\underline{i}kl}$, plus a  contribution that may have information about four-partite entanglement  given by $%
\tau _{ij(kl)}$. Direct calculation shows that for the state (\ref{evol}) $%
\tau _{S_1E_1(S_2E_2)}=\tau _{S_2E_2(S_1E_1)}=0$ for all $p,$ so that $R_{i}=\tau _{%
\underline{i}kl}.$ In Fig. \ref{fig:results2}, we can see the experimental results
corresponding to $\tau _{\underline{i}kl}$  of Eq. (\ref{3partite}) taking the values of $C_{i|jkl}^2$ in the quasipure (QP)
approximation \cite{mintert05b}. It is observed that within the error bars, $%
\tau _{\underline{S_1}S_2E_2}$ (red circles) and $\tau _{\underline{S_2}S_1E_1}$ (black
diamonds) are equal. The same goes for $\tau _{%
\underline{E_1}S_2E_2}$ (blue squares) and $\tau _{\underline{E_2}S_1E_1}$ (magenta
stars). For pure states,  direct calculation of the 3-tangles  can be done using  Eq. (\ref{3partite}). 
We can show that, when the interaction between $S_i$ and $E_i$
is through an AD channel, $\tau _{\underline{S_1}S_2E_2}=\tau _{\underline{S_2}S_1E_1}$ and
$\tau _{\underline{E_1}S_2E_2}=\tau _{\underline{E_2}S_1E_1}$. The curves in Fig.  \ref{fig:results2}
are the direct calculations of (\ref{3partite}) for  the states in Eq. (\ref{evol}).
We can see that there is a qualitative agreement between the data and the theoretical predictions for pure states. It is also  observed that
in all maxima, the experimental states have smaller $\tau _{\underline{i}kl}$ than the prediction for pure states. This is related to the
impurity  of the experimental states and to the fact that $\tau _{\underline{i}kl}$ is calculated in the QP approximation, which is a lower bound
for this quantity \cite{mintert05b}. Notice that $\tau _{\text{S}_{\underline{1}}\text{S}_2\text{E}_2}$ reaches its maximum nearly at $p_{ESD}$, and
analogously, $\tau _{\underline{\text{S}_1}\text{S}_2\text{E}_2}$ reaches its maximum nearly at $p_{ESB}
$. Thus, we conclude that the occurrence of ESD and ESB is related to the
maximum multipartite entanglement, as has been shown in other situations \cite{mcnemes}.

\begin{figure}[tbp]
\centering 
\includegraphics[width=8 cm]{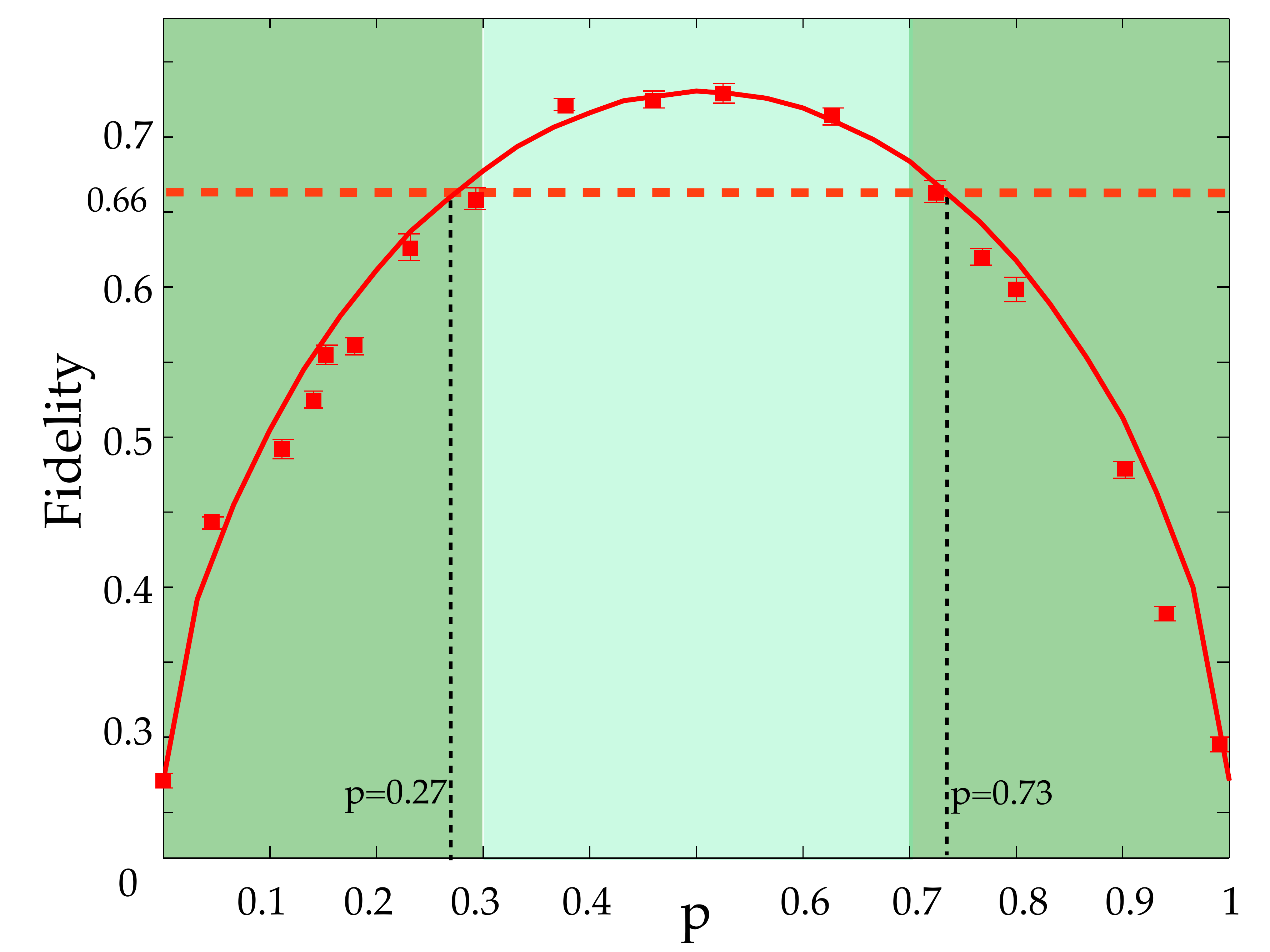}
\caption{Fidelity with respect to the Dick state $\left|{\mathcal{D}}%
\right\rangle $ as a function of p. The four-partite genuine entanglement is
found in the interval $p\in[0.27,0.73]$. Below 0.66  we cannot
guarantee genuine entanglement. The line is the theoretical evolution of the
experimental initial state. }
\label{fig:results3}
\end{figure}

In the previous analysis we learned that the evolved state possesses
tripartite entanglement in the form $\tau _{\underline{i}kl}$. However, we
do not have information about the genuine four-partite entanglement.
For this purpose, we use the fidelities $F_{\left\vert {\psi }\right\rangle }=\left\langle {\psi }%
\right\vert \rho \left\vert {\psi }\right\rangle $ of an experimental state $%
\rho $ with respect to a genuine multipartite entangled state $\left\vert {%
\psi }\right\rangle $, as witnesses of multipartite entanglement.  It has been shown that $\rho $ is genuinely four-partite entangled if  
 $F_{\left\vert {\psi }\right\rangle }>\mathcal{O}$, where $\mathcal{O}$ is
the maximal overlap between $\left\vert {\psi }\right\rangle $ and all the
biseparable states \cite{acin01,wieczorek08}. We note that  the states in Eq. (\ref{evol}) look very similar to the state $%
\left\vert {\mathcal{D}}\right\rangle =1/\sqrt{6}( \left\vert {0000}%
\right\rangle +\left\vert {1111}\right\rangle +\left\vert {0011}%
\right\rangle +\left\vert {1100}\right\rangle +\left\vert {0110}%
\right\rangle$ $ +\left\vert {1001}\right\rangle ) $, which is a Dicke
state with the second and fourth qubits flipped \cite{dicke}.
Moreover, for the case $\left\vert {\psi }%
\right\rangle =\left\vert {\mathcal{D}}\right\rangle $, it was proved that $%
F_{\left\vert {\mathcal{D}}\right\rangle }>2/3$ is sufficient to witness
genuine four-partite entanglement \cite{wieczorek08}. The fidelities $F_{\left\vert {\mathcal{D}}\right\rangle }$ 
for our experimental states
are shown in Fig. \ref{fig:results3}. One can observe that the experimental
points (red squares) are in good agreement with the fidelities between $\left\vert {%
\mathcal{D}}\right\rangle $ and the predictions given by the theoretical
evolution of the initial (experimental) state. We can also see that, at
least in the interval $p\in \lbrack 0.27,0.73]$, the fidelities for the
experimental states are greater than $2/3$, demonstrating the presence of
genuine four-partite entanglement.


\textit{Conclusions:} We presented an experimental investigation of the spread of entanglement from two entangled qubits to their local environments, which
 is quite challenging for other types of physical systems, since in general the
environmental degrees of freedom are unaccessible. We observed the
transition bipartite $\leftrightarrow $ multipartite entanglement along the
evolution and showed that  sudden death of entanglement occurs when all the entanglement becomes
completely multipartite, whereas the sudden birth of entanglement occurs when the entanglement ceases to
be completely multipartite and gets redistributed also in bipartite form. 
We believe that this is the first experimental demonstration of entanglement sudden birth. 
We also presented a novel decomposition of the residual entanglement measures that
allowed us to analyze our results in terms of well-identified 3- and 4-
partite entanglement contributions. In addition, we used the fidelity as a
witness of multipartite entanglement to demonstrate that there is an
interval in the evolution in which the complete system is found in a genuine
four-partite entangled state. Our results represent a
significant step towards a deeper understanding of decoherence processes and the
distribution of entanglement in multi-qubit systems.

\begin{acknowledgements}
Financial support was provided by Brazilian agencies FAPERJ, CNPq, 
CAPES, and the National Institute of Science and Technology for Quantum Information. 
AVH acknowledges financial support from  Programa Ci\^encia sem Fronteiras (CNPq, CAPES).
\end{acknowledgements}

\end{document}